\documentclass[notitlepage]{revtex4-1}
\pdfoutput=1
\usepackage{physics}
\usepackage{graphicx}
\usepackage{dcolumn}
\usepackage{bm}
\usepackage{color}
\usepackage{amsmath}

\begin{document}

\title{Temperature as a quantum observable}
\author{Sushrut Ghonge}
\affiliation{
 Department of Physics, University of Notre Dame, Notre Dame, IN, USA}

\author{Dervis Can Vural}
\email{dvural@nd.edu}
\affiliation{
 Department of Physics, University of Notre Dame, Notre Dame, IN, USA}
 
\date{\today}

\begin{abstract}
In this article, we address the problem of how temperature of a quantum system is observed. By proposing a thought experiment, we argue that temperature must be conceived as an operator and its measurement must necessarily accompany a collapse in the wavefunction. We model a temperature measurement device and determine the expectation value and quantum uncertainty of its readout. Lastly, we explore the consequences of this point of view and propose an experiment to verify if temperature is indeed a quantum observable. 
\end{abstract}

\maketitle
\section{Introduction}
Temperature is the centerpiece of statistical mechanics. Conventionally, it is defined as the change in entropy $S(E)$ with internal energy $E$, at constant volume $V$ and particle number $N$,
\begin{align}\label{def}
\frac{1}{T}=\left.\frac{\partial S(E)}{\partial E}\right|_{V,N}.
\end{align}

Despite its experimental and theoretical practicality, temperature has been the subject of a number of unresolved foundational debates.
Defining entropy and temperature rigorously for systems with discrete energy levels is an open problem \cite{hanggi2016meaning}. For systems with sufficiently closely spaced energy levels, entropy is defined in terms of the density operator $\hat{\rho}$ as $S=\Tr(\hat{\rho} \log \hat{\rho})$ and temperature as $T^{-1}=\Delta S/\Delta E$. Another issue revolves around which of the two definitions of temperature, Boltzmann or Gibbs, is correct \cite{swendsen2017resolving,hanggi2016meaning,swendsen2016negative,swendsen2015gibbs,buonsante2016dispute,dunkel2014consistent,frenkel2015gibbs,cerino2015consistent}. This leads to a debate on whether negative temperatures have physical meaning, since the latter cannot be negative. The question of how temperature should transform under Lorentz transformations has been unclear \cite{sewell2008question,landsberg1996laying,clauser1969proposed}, and is still not resolved satisfactorily \cite{marevs2010relativistic}. A review of definitions of temperature, thermodynamics of small systems, and a discussion on negative temperatures can be found in \cite{puglisi2017temperature}.

The applicability of concepts of thermodynamics, including definitions of temperature, for finite particle systems has also been in question \cite{hartmann2004existence,hartmann2005nano,hartmann2005measurable,hartmann2004local,puglisi2017temperature}. The statistical errors resulting from trying to infer temperature from a finite number of energy measurements was investigated in \cite{mahlerqtp,jahnke2011operational}. The problem of measuring the temperature of non-standard classical systems by real thermometers is studied in detail in \cite{baldovin2017thermometers}. As experimental techniques for measuring temperature of mesoscopic systems improve, a better foundational construction of temperature becomes essential \cite{brites2012thermometry,horodecki2013fundamental,ferraro2012intensive}.


In this article, we formulate how measurements of thermodynamic quantities associated with a quantum systems, particularly temperature, should be formalized. We start by describing an EPR-like paradox for temperature to motivate an alternative definition of temperature in order to resolve the paradox. Specifically, we model what happens when the temperature of a quantum system is measured with a ``physical thermometer''. Our key argument is that temperature cannot be a local realistic variable, and must be redefined as a quantum operator.

Standard quantum mechanics does allow certain quantities to be local-realistic parameters. Examples include mass, charge and spin coupling coefficient. There is no fundamental limit that prohibits monitoring these quantities and knowing their values at all times, with arbitrarily small uncertainty. Furthermore, this knowledge comes at no cost of disrupting other physical observables or wavefunctions. Since these quantities are treated as parameters in wavefunctions and density matrices, and are not operators, they have a single definite value, and no possibility of collapsing into a multitude of eigenvalues. In standard quantum statistical mechanics, temperature is also treated as a parameter, as in (\ref{def}).

The present study is motivated by the  observation that, in practice, all thermal measurement devices operate by mapping temperature to other physical observables such as the length of a mercury column, or the current that passes through a thermoelectric material. If length and current are quantum operators with eigenvalues, eigenstates, expectation values and uncertainties, then it seems reasonable to demand that temperature should also be associated with a quantum operator.

We argue here that temperature must be viewed as an operator rather than as a local realistic parameter. Specifically, we propose a thought experiment that leads to an apparent paradox if one is allowed full knowledge of temperature at all times, without restrictions pertaining operator algebra and wavefunction collapse. We then propose a new definition that treats temperature both experimentally and theoretically as a quantum mechanical operator, $\hat{T}$, and show that this indeed is one way to resolve the paradox. We then discuss a model for a practical quantum thermometer that works by indicating the temperature in terms of position, conceptually similar to the mercury thermometer.  Finally, we discuss possible experimental consequences of the non-standard definition proposed here. 

\section{Thought experiment: Entangling Temperature}

In the energy basis, the density operator $(\hat{\rho})$ for a system in a microcanonical ensemble is a diagonal operator $(\Gamma)^{-1}I$, where $\Gamma$ is the total number of microstates and $I$ is the identity operator. For a system in thermal equilibrium with a heat bath at temperature $T$, it is given by $\hat{\rho}=e^{-\hat{H}/k_B T}/\mathrm{Tr}(e^{-\hat{H}/k_B T})$.
The canonical and grand canonical density operators are also diagonal in the energy basis. 

To ensure that the density operator for a system in a microcanonical ensemble is diagonal in all bases, the postulate of random \textit{a-priori} phases is necessary \cite{Pathria,tolman1938principles}. This postulate states that if a system is in thermal equilibrium, then the wavefunction is an incoherent superposition of the basis states, that is, the probability amplitudes have random \textit{a-priori} phases. This way, a large system in thermal equilibrium always has a diagonal density matrix. An incoherent superposition of energy eigenfunctions cannot be distinguished from a statistical mixture of the same, and the density operators obtained in both cases are equal. The lack of information of energy in one case is equivalent to the lack of information of the phases in another.

We illustrate the difficulty with the conventional definition of temperature with the following thought experiment (shown in figure 1). Consider a collection of $2N$ spins with magnetic moments $\mu$ each, in an external magnetic field $\vec{B}=B\hat{z}$. 

\begin{figure}
\includegraphics[width=\linewidth]{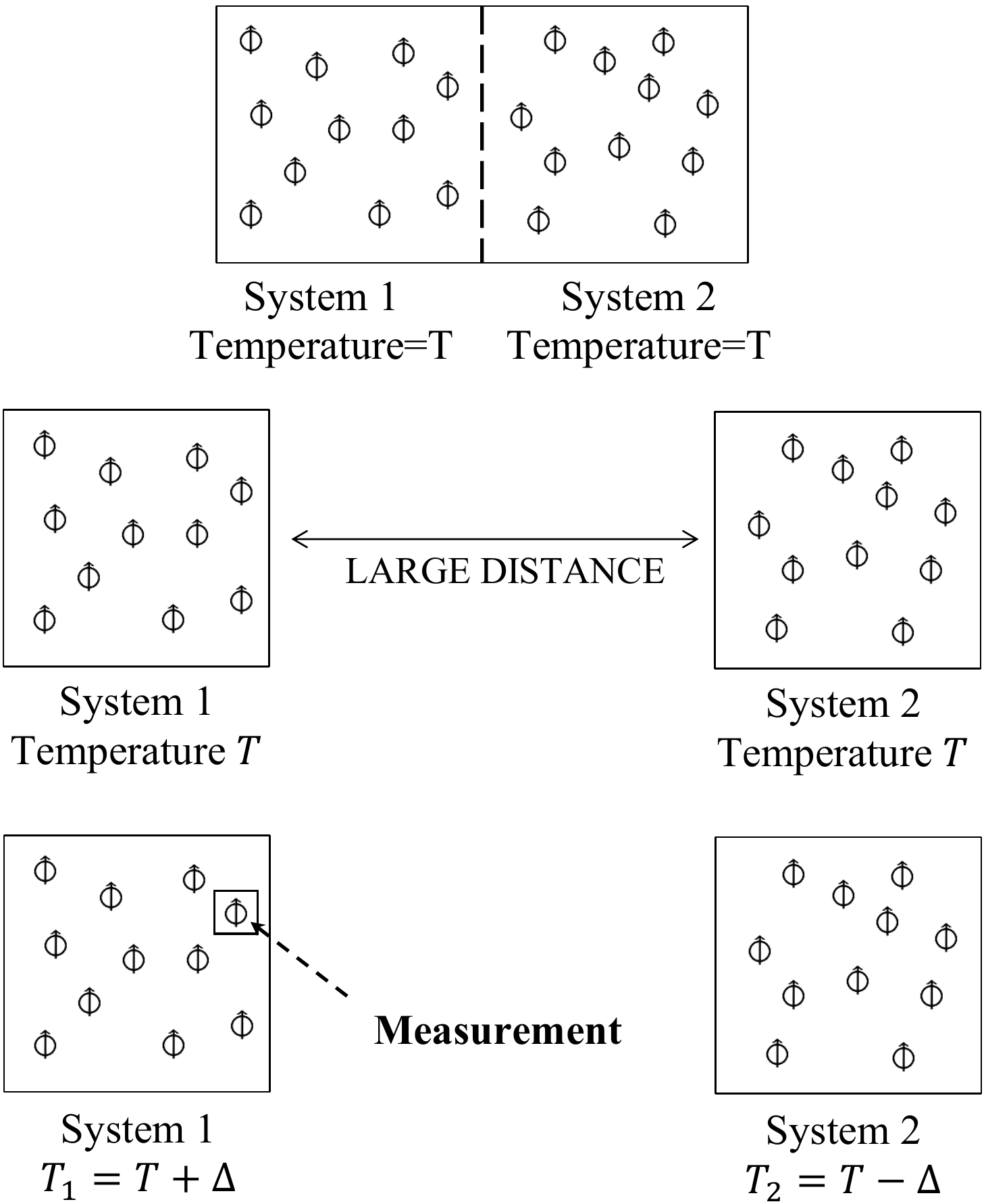}
\caption{\textbf{The thermal analogue of the EPR system.} A system at thermal equilibrium whose particles are entangled through interactions, is split into two subsystems. The subsystems are taken far away and the energy of one particle in one of the subsystems is measured. The measurement causes the subsystem to thermalize to a higher (or lower) temperature. Entanglement causes the other subsystem to thermalize to a lower (or higher) temperature, making it possible to determine that a measurement was made far away.}
\end{figure}

Let the system be completely isolated \footnote{It is possible to sufficiently isolate two state systems from the environment for short durations. For example, see \cite{saeedi2013room} and references therein.} and be prepared in an energy eigenstate with energy $E$. At thermal equilibrium, this system can be described by a microcanonical ensemble with energy $E$.
According to the postulate of random {\it a-priori} phases, the system should be described by the incoherent superposition of all states with energy $E$,
\begin{align}\label{initialpsi}
\hspace{-0.15in}\ket{\psi}=\frac{1}{\sqrt{R}}
\sum_{r=1}^{R}
e^{i\phi_{r}}\ket{\psi_M^{(r)}}, \qquad R=\binom{2N}{M},
\end{align}
where $R$ is the total number of microstates, $\phi_{j}$ are uniformly distributed independent random variables on $[0,2\pi)$, and $|\psi_M^{(r)}\rangle$ is the $r^\mathrm{th}$ wavefunction with $M$ excited spins. For example, $|\psi_{2}^{(1)}\rangle=|11000\ldots\rangle$, $|\psi_{2}^{(2)}\rangle=|10100\ldots\rangle$, and so on.

According to definition (\ref{def}), the temperature is
\begin{align}\label{tspin}
T_{M,2N}= \alpha/\log(2N/M-1), \qquad \alpha=2\mu B/k_B.
\end{align}

Now let us view this system as two subsystems containing $N$ spins each, and suppose that the coupling between subsystems is so weak that the many-body wavefunction will not change as the subsystems are separated apart. Once separated, we measure the state of one of the spins in the first subsystem. This will cause a partial collapse in the many-body wavefunction. After the measurement, the system is allowed to thermalize \cite{srednicki1994chaos,gogolin2016equilibration,malabarba2014quantum}. If the spin is observed in its ground state, the wavefunction becomes, 
\begin{align}\label{finalpsi0}
\hspace{-0.15in}\ket{\psi_{0}}=\frac{1}{\sqrt{R_{0}}}\sum_{r=0}^{R_{0}} e^{i\phi_{0,r}}\ket{\psi_{0,M}^{(r)}}, \qquad R_{0}=\binom{2N\!-\!1}{M}
\end{align}
where $|\psi_{0,M}^{(r)}\rangle$ denotes the $r^\mathrm{th}$ wavefunction with $M$ excited spins, and the first is unexcited. 
On the other hand, if the spin is observed in the excited state, the wavefunction becomes,
\begin{align}\label{finalpsi1}
\hspace{-0.15in}\ket{\psi_{1}}=\frac{1}{\sqrt{R_{1}}}\sum_{r=0}^{R_{1}} e^{i\phi_{1,r}}\ket{\psi_{1,M}^{(r)}}, \qquad R_{1}=\binom{2N\!-\!1}{M\!-\!1},
\end{align}
where, $|\psi_{1,M}^{(r)}\rangle$ denotes the $r^\mathrm{th}$ wavefunction with $M$ excited spins and the first is excited. 

When the single spin in the first subsystem is measured in the ground/excited state, the second system will re-thermalize to a temperature that is higher/lower. The differences in temperature will be
\begin{align}\label{temdif}
\Delta T_0&=T_{M, 2N\!-1}-T_{M,2N}\approx \frac{\alpha}{N\log^2(2N/M)},\\
\Delta T_1&=T_{M-1, 2N\!-1}-T_{M,2N}\approx\frac{-\alpha}{M \log^2(2N/M)}
\end{align}
respectively, where the approximations hold for $2N\gg M \gg1$. These temperature differences vanish in the thermodynamic limit, but in principle, are detectable for finite systems.

If temperature is a local realistic parameter, and not an operator, there is nothing in the formal structure of quantum mechanics prohibiting us from knowing the temperature of either system at all times, without having to collapse any wavefunction. 

As such, an observer detecting a slight increase or decrease in subsystem two would infer that a spin has been measured far away, in subsystem one. Since this can be used for superluminal communication, we must conclude that temperature cannot be a parameter that can be known without a wavefunction collapse. There must be quantum mechanical constraints on its theoretical definition and its experimental measurement. 

The thought experiment presented is essentially a simple thermal analogue of the classical EPR setup \cite{einstein1935can}: We have split a thermal system into two in a way such that their temperatures are entangled; and then perturbed the temperature in one system by collapsing a single spin, to influence the temperature in the other. 

The problem of superluminal communication in the EPR paradox is easily resolved thanks to well-defined spin operators and a quantum measurement process that prohibits non-local influences to measurement statistics \cite{bell2004speakable}. However the same operator formulation is not available to us for temperature and other thermodynamic variables. 

\section{Resolution to the paradox}



To offer a resolution we define a ``temperature operator" $\hat{T}$ and temperature eigenstates to which temperature measurements collapse. In this representation, the states corresponding to particular temperatures must be pure wavefunctions instead of statistical mixtures. We must then show that under these definitions, the measurement statistics cannot indeed be influenced non-locally.


For isolated systems there is a one-to-one correspondence between temperature and total energy $(T=f(E))$ \cite{falcioni2011estimate,jahnke2011operational}, \footnote{In some atypical systems, temperature is not in one to one correspondence with energy. We exclude such systems from this discussion}. Thus, one natural way of constructing $\hat{T}$ is to map temperature to total energy so that temperature eigenfunctions are identical to many-body energy eigenfunctions, and its eigenvalues are the values of temperature corresponding to many-body energy eigenvalues.

For experimental compatibility with the standard statistical mechanics, we define the temperature operator as
\begin{align}\label{tempdef}
\hat{T}=\sum_n f(E_n)\ket{E_n}\bra{E_n}=f(\hat{H})
\end{align}
where $E_n$ are many-body energy eigenvalues, and $f(E)$ is obtained from the mapping between average energy and temperature
\begin{align}\label{tempenergy}
E = f^{-1}(\tau)=\frac{\sum_n E_n \exp(-E_n/k_B\tau)}{\sum_n \exp(-E_n/k_B\tau)}.
\end{align}

How does this operator resolve the paradox?  According to our proposed framework, (\ref{initialpsi}),(\ref{finalpsi0}) and (\ref{finalpsi1}) are all temperature eigenstates corresponding to different temperatures. In order to know the temperature of the second subsystem, one would have to perform a quantum measurement of temperature, which collapses the wavefunction to a ``temperature eigenstate''. Before this measurement, the system is in an incoherent superposition of states that correspond to different numbers of spins in the excited state in the two subsystems. Thus, a temperature measurement on the second subsystem can result in many different values with different probabilities. Although the measurement done on the first subsystem causes a superluminal heat transfer, we will show that this makes no difference in the probability of finding the second subsystem at a certain temperature.

Before the measurement is performed on a spin in the first subsystem, the probability of finding the temperature of the second subsystem to be $T_{m,N}$, i.e. collapsing it to the temperature eigenstate $\ket{T_{m,N}}$, is

\begin{align}\label{sys2}
P^{I}_{m,N}=|\psi(T_{m,N})|^2=\left|\bra{T_{m,N}}\ket{\psi}\right|^{2}&=\frac{1}{R} \sum_{r=1}^{R} \abs{\bra{T_{m,N}}\ket{\psi_M^{(r)}}}^{2}
\end{align}

All states in (\ref{initialpsi}) have $M$ excited spins. Thus, a given term in (\ref{sys2}) is $1$ if the $\psi_M^{(r)}$ corresponds to $m$ excited spins in the second subsystem and $0$ otherwise. The number of wavefunctions with $m$ excited spins in the second box is
$w=\binom{N}{m} \binom{N}{M-m}\nonumber$. Substituting $R=\binom{2N}{M}$ in (\ref{sys2}),
\begin{align}\label{pi}
P^{I}_{m,N}=\binom{2N}{M}^{-1}\binom{N}{m}\binom{N}{M-m}.
\end{align}

When the spin in the first system is measured, it can be either in the ground or the excited state and the wavefunction collapses into (\ref{finalpsi0}) or (\ref{finalpsi1}), with probabilities
\begin{align}\label{probabilities2}
p_0&=e^{-\beta \mu B}/Z = M/2N, \nonumber \\
p_1&=e^{\beta \mu B}/Z = (2N-M)/2N.
\end{align}
If the temperature of the second subsystem is measured after this, the probability $P^{F}_{m,N}$ of finding its temperature to be $T_{m,N}$ is
\begin{align}
P^{F}_{m,N}=\frac{p_0}{R_0}\sum_{r=1}^{R_0} \abs{\bra{T_{m,N}}\ket{\psi_{0,M}^{(r)}}}^{2} + \frac{p_1}{R_1} \sum_{r=1}^{R_1} \abs{\bra{T_{m,N}}\ket{\psi_{1,M}^{(r)}}}^{2}\nonumber
\end{align}
Substituting (\ref{probabilities2}) and using similar counting procedures,
\begin{widetext}
\begin{align}
P^{F}_{m,N}=\frac{M}{2N}{\binom{2N-1}{M}}^{-1} \binom{N}{m} \binom{N-1}{M-m}+\frac{2N-M}{2N}{\binom{2N-1}{M-1}}^{-1} \binom{N}{m} \binom{N-1}{M-m-1}=\binom{2N}{M}^{-1}\binom{N}{m}\binom{N}{M-m} \nonumber
\end{align}
\end{widetext}
which, as we see, is equal to the $P^{I}_{m,N}$ in (\ref{pi}). In other words, while the measurement of a spin in the first subsystem can affect the outcome of a temperature measurement in the second subsystem and lead to superluminal heat transfer, this does not cause any measurable non-local statistical difference.

So far we have described the temperature operator for isolated systems. In general, a system can interact with the environment or with a heat bath.
If temperature is defined in terms of energy, as in (\ref{tempdef}), for a system with finite number of particles, interactions with the environment will cause the wavefunction $|\psi\rangle$ of the system to be in a superposition of different energies and hence a superposition of temperatures. A temperature measurement as defined in eq. (\ref{tempdef}), could then yield a number of different values, with probability $|\psi(T)|^2=|\langle T|\psi\rangle|^2$,  where $|T\rangle$ is an eigenstate of the operator eqn (\ref{tempdef}). Note that if the system in the canonical ensemble has an infinite number of particles, it is equivalent to a microcanonical ensemble and its energy is constant. Therefore, as in the case of isolated systems, temperature measurements always give the same value. From the point of view presented in this paper, such system remains in a temperature eigenstate, with zero quantum uncertainty in temperature.

\section{Alternative Temperatures for Alternative Thermometers} 

While defining a temperature operator by mapping it to the Hamiltonian does provide a solution to the ``thermal EPR paradox'' discussed above, this mapping is not unique. Other variables such as kinetic energy, magnetization or position can also be mapped to temperature. These alternative operators will have fundamentally different properties and need not even commute with each other.

A general temperature measurement involves a system $(S)$ and a thermometer $(M)$ that interacts with the system. The general Hamiltonian ($\hat{H}$) for the system-thermometer supersystem is $\hat{H}=\hat{H}_S+\hat{H}_M+\hat{H}_{\mathrm{int}}$. We will denote corresponding temperatures, energies and wavefunctions with the same subscripts, $S$ and $M$.

When a thermometer is coupled to the system, the supersystem thermalizes to a temperature eigenstate with temperature $T_S$. The thermometer must have an indicating (many-body) variable $\hat{X}$, e.g. energy, position or magnetization, the expectation value of which depends sensitively on temperature, $\langle X_S\rangle=f^{-1}(T_S)$, analogous to (\ref{tempenergy}). 

For the thermometer to read out the instantaneous temperature at a given time, it must perform a single quantum measurement of $\hat{X}$, and then map it to a temperature. Once the measurement is complete, the supersystem collapses into a temperature eigenstate $T_m=f(X_m)$. As a concrete example, in the next section we study the properties of a simple thermometer that utilizes position as an indicating variable.

\section{Position Thermometer}
A standard mercury thermometer displays the temperature in terms of the amount of expansion of a mercury column. Thus, from an empirical and operational point of view, if position is an operator, then so should temperature.

Conceptually motivated by the mercury column, we now model a ``device'' that displays temperature in terms of the position of a ``indicator needle''.

The device simply consists of a collection of $N$ one-dimensional harmonic oscillators with natural frequency $\omega$ and mass $m$. We will take the indicating variable to be the sum of their squared positions, $\hat{Y}=\frac{1}{N} \sum_i \hat{x}^2_i$. This is an alternative convention to (\ref{tempdef}), where now temperature and position commute, and a measurement of an ensemble of positions will result in a collapse of the wavefunction of the system to a temperature eigenstate $|T\rangle$.

A full description of out-of-equilibrium dynamics of the thermometer would require either solving the Schrodinger's equation for the entire supersystem or describing the system as a source of dissipation and noise \cite{weiss2012quantum,gardiner2004quantum}. For the sake of simplicity, we will instead assume that the system is isolated, at thermal equilibrium and in a temperature eigenstate $T_S$. We also assume that the system is much larger than the thermometer, so that the system-thermometer supersystem thermalizes to a temperature almost exactly equal to $T_S$.

\begin{figure}[b]
\includegraphics[width=\linewidth]{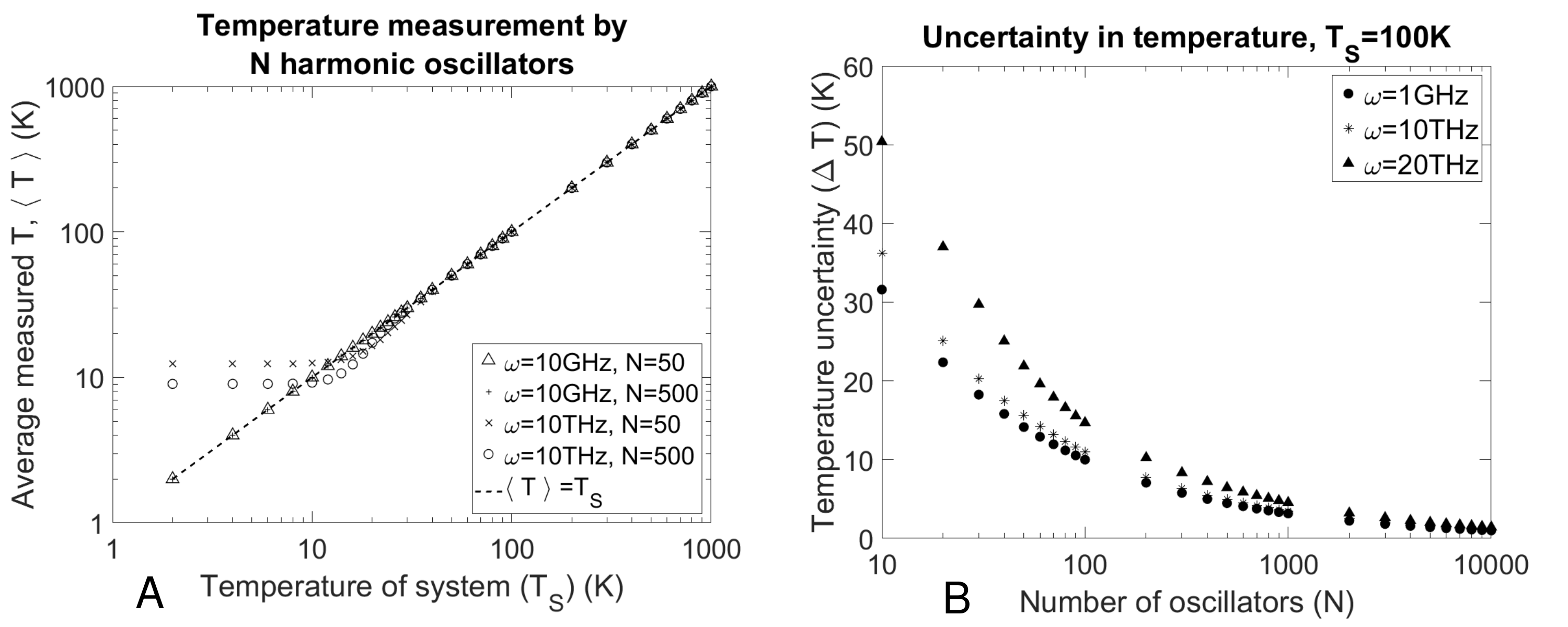}
\caption{The system has a temperature $T_S$ and the thermometer consists of $N$ harmonic oscillators of a given natural frequency $(\omega)$ and the mass $m$ of each harmonic oscillator is assumed to be equal to the mass of a typical molecule (6 a.u.). A. The expectation value of the temperature operator is plotted against the temperature of the system for different $T_S$, $\omega$ and $N$. Oscillators with very high natural frequency are inaccurate at low temperatures, but their accuracy increases with $N$. B. The uncertainty in temperature is plotted against $N$ for various $\omega$. As $N$ increases, the uncertainty in temperature decreases. For $\omega<10$THz, the uncertainty is not strongly dependent on $\omega$.}
\end{figure}

When the thermometer reaches equilibrium with the system, its wavefunction and density operator is fully defined by the (non-operator) quantity $T_S$, the temperature of the system,
\begin{align}\label{ho-psi}
\ket{\psi_M}=\frac{1}{\Gamma}\bigg[\sum_{j=1}^{\infty}\ket{E_j}\exp(-E_j/2k_BT_S)e^{i\phi_j} \bigg],\\
\hat{\rho}_M=\exp(-\hat{H}_M/k_BT_S)/Z \ ; \qquad Z=\Tr(\exp(-\hat{H}_M/k_BT_S)) \nonumber
\end{align}

where $\Gamma$ is a normalization constant and $\phi_j$ are random phases that occur along with the Boltzmann factors due to tracing over the system degrees of freedom. 

Given that the system is in a temperature eigenstate $T_S$ what is the quantum expectation value and quantum uncertainty of $\hat{T}$ for the ``indicator needle'' of our temperature measurement device? 

At equilibrium with the system, the variance in position is (see Appendix),
\begin{align}\label{single-var}
\langle \hat{x}^2 \rangle = \frac{\hbar}{2m\omega}\coth({\frac{\hbar\omega}{2k_B T_{\text{S}}}}).
\end{align}

We substitute the quantum variable $\hat{Y}$ for $\langle x^2\rangle$ and $\hat{T}$ for $T_S$, so that the temperature reading of the thermometer is mapped to a physical variable,
\begin{align}\label{calibration}
\hat{T}=\frac{\hbar\omega}{2k_B} \Big[\text{arcoth}(2m\omega \hat{Y}/\hbar)\Big]^{-1}.
\end{align}
This equation can be viewed as a calibration curve for our thermometer, in the sense that once an observer measures $\hat{Y}$, obtains an eigenvalue $Y_m$, she would use (\ref{calibration}) to get the instantaneous temperature $T_m$, an eigenvalue of $\hat{T}$.


The position distribution for a single thermalized harmonic oscillator is (see Appendix)
\begin{align}
P(x)=\exp(-x^2/2\sigma^2)/\sqrt{2\pi\sigma^2}
\end{align}
where $\sigma^2=(\hbar/2m\omega)\coth(\hbar\omega/2k_B T_S)$. Therefore the probability distribution for $y=x^2$ is given by
\begin{align}\label{single-ho-2}
P(y)= \frac{\mathrm{d}x}{\mathrm{d}y}\bigg|_{x}P(x)+\frac{\mathrm{d}x}{\mathrm{d}y}\bigg|_{-x}P(-x) = \frac{1}{\sigma\sqrt{2\pi y}}\exp(-\frac{y}{2\sigma^2}).
\end{align}

From the above distribution we determine that $\langle y \rangle=\sigma^2$ and $\mbox{Var}(y)=2\sigma^4$. We can use the central limit theorem to approximate the distribution for the many-body variable $Y_m$,
\begin{align}\label{many-ho}
P(Y_m) = \abs{\bra{Y_m}\ket{\psi_M}}^2 \approx\sqrt{\frac{N}{2\pi\sigma^4}}\exp(-\frac{N(Y_m-\sigma^2)^2}{2\sigma^4})
\end{align}

Finally, the probability distribution for temperature is obtained from (\ref{calibration}),
\begin{align}\label{ho-temp}
\mathcal{P}(T) = \abs{\bra{T}\ket{\psi_M}}^2 &=\frac{\mathrm{d}Y_m}{\mathrm{d}T}P(Y_m)=\frac{\hbar^2}{4mk_BT^2}\mathrm{csch}^2\bigg(\frac{\hbar\omega}{2k_B T} \bigg)P(Y_m(T,T_S))
\end{align}
where
\begin{align}
P(Y_m(T,T_S)) &=\sqrt{\frac{N}{2\pi\sigma^4}}\exp\left\{\frac{-N\hbar^2}{8m^2\omega^2\sigma^4}\left[\coth(\frac{\hbar\omega}{2k_B T})-\coth(\frac{\hbar\omega}{2k_B T_S})\right]^2\right\}.
\end{align}

The average temperature measured by this thermometer is the expectation value of the temperature operator and is approximately equal to the temperature of the system. Fig. 2. shows the expectation value and quantum uncertainty associated with the temperature operator for various $T_S$, $N$ and $\omega$.

Equation (\ref{tempenergy}) is correct only when the energy on the left side is the expectation value of energy (or another indicating variable). Therefore, to measure the exact temperature, one must make repeated measurements of the variable, take their average and solve eq. \ref{tempenergy} (or a similar equation for that variable). However, quantum mechanics only allows instantaneous measurements defined by linear operators. As such, we cannot define an operator as the average of several measurements. Instead, we define the temperature operator in terms of the measured energy or position. Therefore, the expectation value of the temperature operator may not be equal to the temperature of the system, which is equal to the temperature eigenvalue corresponding to the average value of the variable. This deviation can be seen in Fig. 2 for oscillators with high natural frequencies. This problem goes away when the number of particles is so large that the wavefunction of the thermometer is approximately a delta function at the expectation value.

\section{Experimental Consequences}
The operator view of temperature has testable consequences. Specifically, it should be possible to construct systems whose wavefunctions are superpositions of two or more different temperature eigenfunctions. We propose an experiment to observe such temperature superpositions in Fig.3. A paramagnetic material is kept at a temperature $T_1$. A photon is passed through a beam splitter towards the material, which if absorbed, will raise the temperature of the material to $T_2$. This causes the material to be in a superposition of two temperature states, and hence two magnetizations. A spin source placed near the material emits pulses of particles with identical spin state. When the spin pulse reaches the screen with two slits, the spins will be in a superposition of two different scattered states. Therefore, we should be able to see an interference pattern on the rear screen. If just a spot is observed behind either of the slits, then this will indicate that the temperature operator that we have described is not the correct way to resolve the paradox discussed above. 

\begin{figure}
\includegraphics[width=\linewidth]{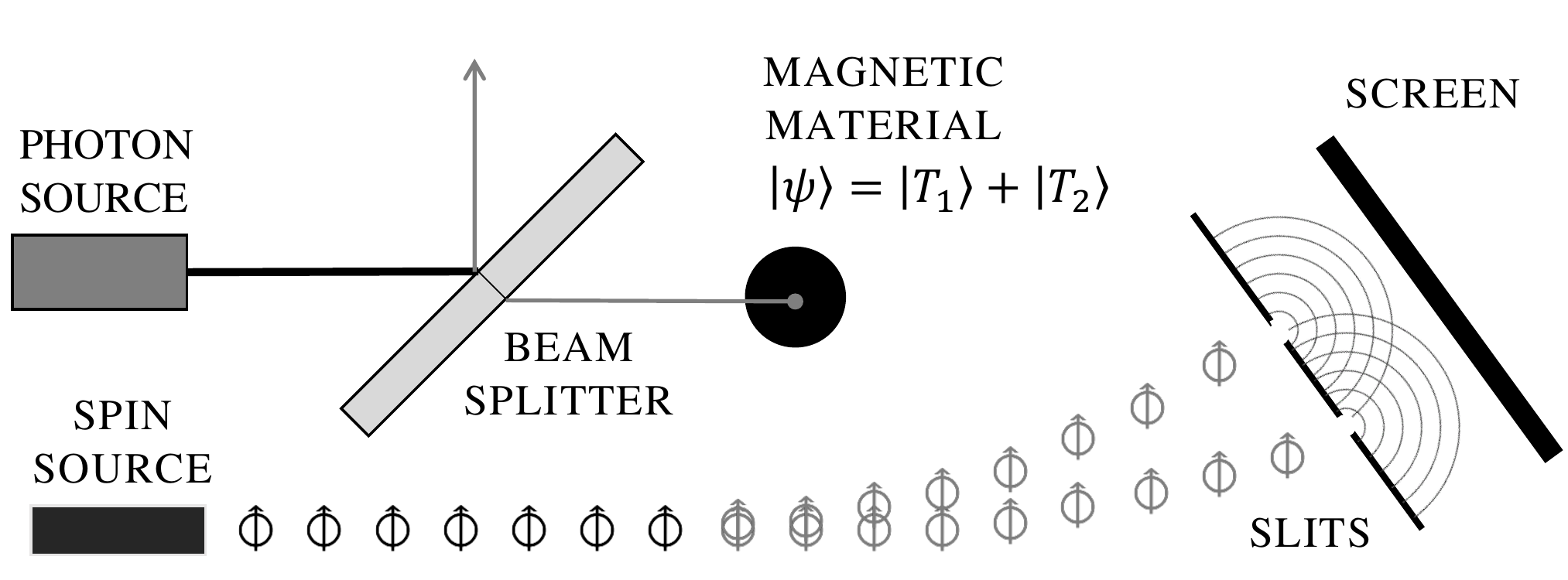}
\caption{{\bf Proposed experiment to observe temperature superposition states.} A paramagnetic substance is kept at a temperature $T_1$. A source emits a photon of high energy, enough to slightly raise the temperature of the substance upon absorption. A beam splitter is kept between the source and the sample so that the sample is in a superposition of two temperature states, and thus two different magnetizations. When the spins emitted by a source reach the screen with two slits, they will be in a superposition of two different scattered states. Therefore, we should be able to see an interference pattern on the rear screen. If just a spot is observed behind either of the slits, then this will indicate that the temperature operator that we have described is not the correct way to resolve the paradox.}
\end{figure}

Furthermore, if the material is ferromagnetic and slightly below its Curie temperature, a photon projected through a beam splitter should bring the system into a superposition of temperature eigenstates that corresponds to a superposition of macroscopic phases. A substance can be cold and hot, or a ferromagnet and a paramagnet at the same time. 

Note that if each spin acquires a different random phase while passing by the thermalized magnetic material, the interference pattern will be washed out. Therefore it is important that the spins are ejected in a sufficiently tight pulse, so that they all experience the same magnetic state of the material. 

\section{Discussion}
We presented a thought experiment that leads to a violation of causality if the standard definition of temperature is used. We then suggested a possible way of resolving the issue by promoting temperature to a quantum operator. Specifically, we considered two possible operator definitions, one based on energy and another based on position. We considered a model thermal measurement device that displays the temperature of a system through the position of an indicator, and determined expectation values and quantum uncertainties associated with the ``indicator needle'' of this device, and thus, that of temperature.

In our framework, the temperature operator must be system specific, in the sense that a quantum thermometer needs to be calibrated according to the system that it measures. For example, for an energy-measuring thermometer, the mapping (\ref{tempenergy}) between energy and temperature depends on the density of states of the system. Thus, a thermometer must be re-calibrated whenever a different system is to be measured. Furthermore, since the density of states will be perturbed upon the coupling of the thermometer and the system, a thermometer needs to be calibrated also according to how strongly it interacts with a system. Similar effects has already been considered for classical thermometers \cite{baldovin2017thermometers}. 

It was pointed out in \cite{gemmer2009observabilityin,*gemmer2009observabilityex,*gemmer2009pressure}, in a very similar spirit to the present work, that temperature and pressure are not fundamental observables, but can only be inferred indirectly. For example, the temperature of a system can be estimated by doing one single (classical) measurement of energy. This procedure and statistical errors resulting from it were studied in \cite{mahlerqtp,jahnke2011operational}. Estimation of the temperature of a system by repeated energy measurements on the same system was discussed in \cite{de2017estimating}. Some eigenstate specific temperatures and statistical inference errors arising from them have been investigated in \cite{borgonovi2017temperature}. 

Even though \cite{borgonovi2017temperature,gemmer2009observabilityin,gemmer2009observabilityex,mahlerqtp,jahnke2011operational,de2017estimating} discuss temperature uncertainties or assign temperatures to eigenstates of other operators, these investigations still view temperature as a local realistic parameter,  in the sense that systems do have a definite temperature, but this temperature cannot be inferred accurately due to the fluctuations in energy and the finite number of measurements. This view of temperature strictly prohibits interference effects. 

In our view however, an uncertainty in temperature can be intrinsically quantum mechanical, as discussed at the end of section III. In most experiments these quantum mechanical uncertainties will not manifest as interference effects due to incoherence between thermal states in thermal superpositions. However we can hope that macroscopic superposition experiments such as that shown in Fig.3 can tease out the quantum nature of temperature.

Recent theoretical studies of nanoscale thermometers such as \cite{miller2018energy} support the idea of a quantum uncertainty for temperature. Future experiments similar to that described in Fig.3 may further help distinguish between alternative resolutions to our EPR-like problem, developing and strengthening the foundations of temperature beyond the elementary arguments offered here. 

Furthermore, what we have argued here for temperature might also make sense for other macroscopic thermodynamic entities such as free energy, entropy, volume and pressure, so that they too are redefined in terms of the Hamiltonian or other operators \cite{borowski2003concept,gemmer2009pressure}.

\bibliography{refTemp}
\appendix*
\section{Derivation of harmonic oscillator position distribution}
For the sake of completeness here we include a derivation of the position distribution of a thermalized harmonic oscillator  \cite{feynman1972statistical,cohen,barragan2018harmonic}.

The Hamiltonian for a one-dimensional harmonic oscillator is 
$$\hat{H}=\hbar \omega (a^{\dagger}a+1/2)$$
So the diagonal element of the density matrix in the position basis $\rho(x)$ can be written as
\begin{align}
\rho(x)=Z^{-1}\exp(-\lambda/2)f(x)\\
\text{where, }\lambda=\beta\hbar\omega \text{ and } f(x)=\bra{x}\exp(-\lambda a^{\dagger}a)\ket{x}
\end{align}

We calculate the variation in $f(x)$ when $x$ is changed slightly:
\begin{align}
\ket{x+\mathrm{d}x}=(1-\frac{ip}{\hbar}\mathrm{d}x)\ket{x} \nonumber \\
\label{dfx} \text{Hence, } f(x+\mathrm{d}x)=f(x)+\frac{i\mathrm{d}x}{\hbar} \bra{x}\big[p,\text{exp}(-\lambda a^{\dagger}a)\big]\ket{x}.
\end{align}

Since $x$ is proportional to $(a+a^{\dagger})$ and $p$ is proportional to $(a-a^{\dagger})$, we evaluate the quantities:
\begin{align}\label{aexp}
a\exp(-\lambda a^{\dagger}a)\ket{\phi_n}=\sqrt{n}\exp(-\lambda n)\ket{\phi_{n-1}}\\
\label{aexp2} \exp(-\lambda a^{\dagger}a)a\ket{\phi_n}=\sqrt{n}\exp(-\lambda (n-1))\ket{\phi_{n-1}}
\end{align}

From \ref{aexp} and \ref{aexp2},
\begin{align}\label{a1}
\exp(-\lambda a^{\dagger}a)a=\exp(\lambda)a\exp(-\lambda a^{\dagger}a)
\end{align}

Similarly,
\begin{align}\label{a2}
\exp(-\lambda a^{\dagger}a)a^{\dagger}=\exp(-\lambda)a^{\dagger}\exp(-\lambda a^{\dagger}a)
\end{align}

Subtracting \ref{a2} from \ref{a1},
\begin{align}
[a-a^{\dagger},\exp(-\lambda a^{\dagger}a)]=\tanh(\lambda/2)[a+a^{\dagger},\exp(-\lambda a^{\dagger}a)]_{+}
\end{align}

Using $x=\sqrt{\hbar/2m\omega}(a+a^{\dagger})$ and $p=\sqrt{\hbar/2m\omega}(a+a^{\dagger})$:
\begin{align}\label{p-exp}
[p,\exp(-\lambda a^{\dagger}a)]=im\omega \tanh(\lambda/2)[x,\exp(-\lambda a^{\dagger}a)]_{+}
\end{align}

Substituting \ref{p-exp} in eq. \ref{dfx}, we get:
\begin{align}
f(x+dx)-f(x)=-\frac{m\omega}{\hbar}\mathrm{d}x \tanh(\lambda/2)\bra{x}[x,\exp(-\lambda a^{\dagger}a)]_{+}\ket{x} \nonumber \\
=-2x\frac{m\omega}{\hbar}\tanh(\lambda/2)f(x)\mathrm{d}x
\end{align}

Therefore, $f(x)$ satisfies the differential equation
\begin{align}\label{fxdiff}
\frac{\mathrm{d}f(x)}{\mathrm{d}x}+\frac{2x}{\xi^2}f(x)=0\\
\text{where, } \xi=\sqrt{\frac{\hbar}{m\omega}\coth(\frac{\lambda}{2})}.\nonumber
\end{align}

The solution to \ref{fxdiff} is
\begin{align}
f(x)=f(0)\exp(-x^2/\xi^2)
\end{align}

Since the probability distribution is normalized to 1, we finally obtain
\begin{align}
\rho(x)=\frac{1}{\xi\sqrt{\pi}}\exp(-x^2/\xi^2)
\end{align}

Therefore, $\langle \hat{x} \rangle =0$ and $\langle \hat{x}^2 \rangle = \frac{\hbar}{2m\omega}\coth(\frac{\beta\hbar\omega}{2}).$
The same result has been derived using other techniques in \cite{feynman1972statistical,barragan2018harmonic}.


\end{document}